\documentclass
	[titlepage,twoside,envcountsame,a4paper,fleqn,oribibl,leqno]
	{llncs}

\usepackage{amssymb}

\newcommand{\xunion}		{\cup}
\newcommand{\xUnion}		{\bigcup}
\newcommand{\xintersect}	{\cap}

\newcommand{\xdiff}		{\backslash}

\renewcommand{\emptyset}	{\varnothing}

\newcommand{\xrange}[4]		{\ensuremath{#1_#2, \ldots, #3_#4}}
\newcommand{\xrangen}[1]	{\ensuremath{#1_1, \ldots, #1_n}}

\newcommand{\xsubst}[2]		{\ensuremath{#1\!\mapsto\! #2}}
\newcommand{\xsubsts}[2]	{\ensuremath{\{#1\!\mapsto\! #2\}}}

\newcommand{\xtupleI}[1]	{\ensuremath{\left\langle #1\right\rangle}}
\newcommand{\xtupleII}[2]	{\ensuremath{\left\langle #1; #2\right\rangle}}

\newcommand{\xstupleII}[2]	{\ensuremath{\left\langle #1\,;\,#2\right\rangle}}

\newcommand{\xtuple}[1]		{\xtupleI{#1}}

\newcommand{\xOrd}[1]		{\ensuremath{\mathcal{O}\left(#1\right)}}

\newcommand{\xeclipse}		{\textup{ECL\textsuperscript{\itshape i}PS\textsuperscript{\itshape e}}}
\newcommand{\xopl}		{\textup{\sffamily OPL}}

\spnewtheorem{consequence}[theorem]{Consequence}{\itshape}{}
\spnewtheorem*{example*}{Example}{\itshape}{}

\newcommand{\xalgorithm}[1]	{{\sffamily\bfseries #1}}
\newcommand{\xalgocomment}[1]	{{\sffamily\small #1}}
\newlength{\xalgokeyword}
\newlength{\xalgotab}
\newlength{\xalgotabinitial}
\newlength{\xalgoskip}
\newenvironment{algorithm}
{%
\begin{tabbing}
\hspace{\xalgotabinitial}\=%
\hspace{\xalgotab}\=\hspace{\xalgotab}\=%
\hspace{\xalgotab}\=\hspace{\xalgotab}\=%
\hspace{\xalgotab}\=\hspace{\xalgotab}\=\kill\>\+}
{\end{tabbing}%
}

\newcommand{\xafor}		{\xalgorithm{for\hspace{\xalgokeyword}}}
\newcommand{\xawhile}		{\xalgorithm{while\hspace{\xalgokeyword}}}
\newcommand{\xaif}		{\xalgorithm{if\hspace{\xalgokeyword}}}
\newcommand{\xathen}		{\xalgorithm{\hspace{\xalgokeyword}then}}
\newcommand{\xaelse}		{\xalgorithm{else\hspace{\xalgokeyword}}}

\newcommand{\setnewlength}[2]	{\newlength{#1}\setlength{#1}{#2}}

\newlength{\xsetlengthtmpstore}
\newcommand{\setlengthtmp}[2]
		{\setlength{\xsetlengthtmpstore}{#1}\setlength{#1}{#2}}
\newcommand{\resetlength}[1]
		{\setlength{#1}{\xsetlengthtmpstore}}

\newcommand{\xC}	{\ensuremath{\mathcal C}}
\newcommand{\xD}	{\ensuremath{\mathcal D}}

\newcommand{\xP}	{\ensuremath{\mathcal P}}
\newcommand{\xQ}	{\ensuremath{\mathcal Q}}

\newcommand{\xRac}	{\mbox{\rmfamily\upshape (ac)}}
\newcommand{\xRarx}	{\mbox{\rmfamily\upshape (arr$_x$)}}
\newcommand{\xRary}	{\mbox{\rmfamily\upshape (arr$_y$)}}
\newcommand{\xRara}	{\mbox{\rmfamily\upshape (arr$_a$)}}
\newcommand{\xRarap}	{\mbox{\rmfamily\upshape (arr$_a'$)}}

\newcommand{\xRSarr}	{\ensuremath{\mathcal R_{\mbox{\upshape\rmfamily arr}}}}

\newcommand{\xcond}[1]	{\ensuremath{\langle\mbox{\itshape #1}\rangle}}

\newcommand{\xarrac}	{{\sffamily arr-ac}}

\title{Constraint Propagation \\in Presence of Arrays}
\author{Sebastian Brand}
\institute{CWI \\ P.O.~Box 94079 \\ NL-1090 GB
	Amsterdam \\ The Netherlands \\
	Sebastian.Brand@cwi.nl\\
	www.cwi.nl/{\textasciitilde}sbrand/}

\begin{document}
\sloppy

\maketitle

\begin{abstract}
We describe the use of array expressions as constraints,
which represents a consequent generalisation of the \texttt{element}
constraint.
Constraint propagation for array constraints is studied theoretically,
and for a set of domain reduction rules the local consistency they enforce,
arc-consistency, is proved.
An efficient algorithm is described
that encapsulates the rule set and so inherits the
capability to enforce arc-consistency from the rules.

\end{abstract}


\section{Introduction}

Many problems can be modelled advantageously using "look up" functionality:
access an object given an index.
Imperative programming languages provide arrays for this.
With \texttt{i} one of \texttt{1}, \texttt{2}, \texttt{3} and
a definition such as \mbox{\texttt{integer a[3]}},
the construct \texttt{a[i]} represents an integer variable,
while with a definition \mbox{\texttt{b[] = \{5, 7, 9\}}}
the `value' of \texttt{i} according to table \texttt{b}
can be looked up by \mbox{\texttt{x = b[i]}}.

A usual condition for look-up expressions to be valid
is that the index be known when the expression is evaluated.
In a constraint programming environment this is a restriction
that can be disposed of.
The binary \texttt{element} constraint (originally in CHIP, \cite{dh:chip}),
semantically equivalent to a lookup expression using a 1-dimensional array,
allows a variable as the index
and a variable for the result, constraining both.
It has proved very beneficial to allow a variable for the index.
Many important problems (scheduling, resource allocation, etc.)
modelled as CSPs make use of this constraint.

\xopl, a modelling language for combinatorial optimisation and constraint
programming (\cite{pvh:opl}), supports arrays of constants and variables,
and indexed by variables (or other expressions).
These array expressions are most general.
However, domain reduction in \xopl\  is weaker than possible,
for instance the reduction for an index variable depends
on its position (\cite{pvh:opl}, p. 100).

In this work we study constraint propagation enforcing arc-consistency for
general array expressions.  Arrays are multidimensional, and variables can
occur wherever constants can.
An expression $x=a[\xrangen y]$ is seen as a constraint on the variables
$x$, and $\xrangen y$, and all variables collected in the array $a$.

\begin{example*}
Consider an application of arrays. 
Assume a conventional crossword grid, with entries for words
in the rows and columns and remaining fields blackened.
Further consider a set of words, a subset of which is to be filled into
the entries in the grid.
A natural formulation of this problem as a CSP
is to take for each word entry a variable $E_i$ 
whose initial domain is the set of words that fit in the entry length-wise.

The words, split up in their letters, are collected in a two-dimensional array $l$
such that $l[w,p]$ represents the letter of word $w$ at position $p$.
The conditions on crossings of entries are
then easily stated as constraints.
A crossing of field $E_1$ at position $4$ 
and field $E_2$ at position $3$ is stated as
\mbox{$l[E_1, 4] = l[E_2, 3]$}.
An additional \texttt{alldifferent} constraint on the $E_i$ ensures
that no two word entries contain the same word.

Enforcing arc-consistency for array expressions
solves some instances of the crossword problem
without any backtracking
(\cite{pvh:cslp}, p. 140, which uses a custom constraint
for crossing entries that is equivalent to the one here).
\end{example*}


\section{Preliminaries}

A constraint satisfaction problem
\xtupleII {\xC}{\xD} consists of
a set of variables (implicit here),
a set \xD\ of domain expressions $x \in D_x$
that associate with every variable a set of admissible values,
and a set \xC\ of constraints.
A constraint is a relation on a set of variables
that is a subset of the cartesian product of their domains.

A \emph{solution} for a constraint
is an assignment of values to its variables
that is consistent with the constraint.
A solution for a CSP is an assignment that is a solution
for all its constraints.
A CSP, or a constraint, is \emph{satisfiable} if a solution exists.
A domain value, or a partial solution,
is \emph{supported} if it is part of a solution.

Local consistency notions, weaker approximations of (global) satisfiability,
are essential in constraint solving.
A central one is arc-consistency (\cite{macworth:cnr}).
We disregard the arity of constraints and define:
a constraint is \emph{arc-consistent},
if all domain values of all its variables are supported.
A CSP is arc-consistent if all its constraints are.

\subsection{A Rule-based Formalism}

Constraint programming can be seen as transforming
a CSP into one or several simpler but equivalent CSPs
in a rule-based way.
This view allows separate consideration
of the reductive strength of some set of
constraint propagation rules, and its scheduling.
The transformations on CSPs lend themselves to a declarative formulation.
We adopt the proof theoretic formalism of \cite{apt:ptv},
and introduce the elements relevant here.

A transformation step from a CSP \xP, the premise,
to a CSP \xQ, the conclusion, by application of a rule (r)
and possibly subject to a side condition \xcond{C\,} on \xP\ 
is represented as
\[
\mbox{\rmfamily (r)}\quad
\frac{\xP}{\xQ}\quad\xcond{C\,}
\]

Two CSPs \xP\ and \xQ\ are \emph{equivalent}
if all variables in \xP\ are present in \xQ\
and every solution for \xP\ 
can be extended to a solution
for \xQ\ by an assignment to variables only in \xQ.
If a rule application preserves equivalence
then the rule is \emph{sound}.

A rule is required to be relevantly \emph{applicable}, that is,
the result \xQ\ must be different from \xP\ in the sense
that the set of domain expressions or the set of constraints
changes.
If a rule, or a set of rules, is not applicable to \xP\ then \xP\ is
\emph{stable} or \emph{closed under} it.
Applying a rule to a constraint means applying it to
the CSP consisting only of this constraint.

\paragraph{Notation.}
Domain expressions $v \in D_v$ used in rules are implicitly
represented in the set \xD.
Replacing a domain expression present in \xD\
is denoted by \mbox{$\xD,v \in D_v^\mathit{new}$}.
If in $\xP = \xtupleII {\xC}{\xD}$ the set of constraints
consists of only one constraint, \mbox{$\xC = \{\mathit{con}\}$}, then we
may just write $\xP = \xtupleII {\mathit{con}}{\xD}$.
The expression \xsubst st denotes
a substitution, assignment, or mapping, from $s$ to $t$.

\begin{example*}
We illustrate these concepts with a rule-based characterisation of
arc-consistency.  A constraint $C$ is arc-consistent
if for all variables $v$ of $C$ and all values $d \in D_v$
an instantiation of $v$ to $d$ in $C$, written $\xC\xsubsts vd$,
retains satisfiability.  If $\xC\xsubsts vd$ is not satisfiable
then $d$ is redundant and can be removed from $D_v$.  The resulting
CSP is equivalent to the original one.
This principle is captured in a rule:
\begin{lemma}\label{thm:acrule}
A satisfiable constraint $C$
is arc-consistent iff it is closed under the application of
\[
\xRac\qquad
\frac
{ \xstupleII {C}{\xD} }
{ \xstupleII {C}{\xD, v \in D_v \xdiff \{d\}} }
\quad
\xC\xsubsts vd \mbox{ has no solution }
\]
\qed
\end{lemma}
\end{example*}


\section{Arc-consistency for Array Constraints} 


An array $a$ of arity $n$ is a set of mappings
\xsubst {\mathit{index}}{\mathit{variable}}.
$\mathit{index}$ is a unique $n$-tuple of constants, 
$\mathit{variable}$ is a variable with a domain.
The array expression $a[\xrangen b]$
evaluates to $v$, if
$a$ contains a mapping \xsubst {(\xrangen b)}v, otherwise it is not defined.
(in what follows it is assumed that indices accessing $a$ are valid).
Note that arrays of constants come as a specialisation of this model.

\subsection{Simple Array Constraints}

Array expressions $a[\xrangen y]$ are functional.  The simplest
extension to a constraint is the equality constraint
$C \equiv \xtupleI {x=a[\xrangen y]}$.
We establish arc-consistency first for this case,
and discuss subsequently compound (nested) array expressions.
Also, occurrences of variables are restricted in that no variable in
the constraint may occur more than once ($C$ is a linear).
Note that the variables of $C$
are $x$, $\xrangen y$, and all variables $v$
for a valid $(\xrangen b)$
and \xsubst {(\xrangen b)}v in $a$.
Such a $v$ will from now on be denoted directly as $a[\xrangen b]$.

\begin{theorem}[Arc-consistency for arrays]
\label{thm:arr}
A satisfiable linear equality constraint \xtupleI {x=a[\xrangen y]}
is arc-consistent iff it is closed under the rule set \xRSarr:
\setlengthtmp{\mathindent}{2mm}
\setnewlength{\xcondindent}{25mm}
\[
\xRarx\qquad
\frac
{ \xstupleII {x=a[\xrangen y]}
	{\xD} }
{ \xstupleII {x=a[\xrange y1yn]}{\xD, x \in D_x \xintersect
	\left(\xUnion_{b_i\in D_{y_i}} D_{a[\xrange b1bn]}\right)} }
\]
\[
\xRary\qquad
\frac
{ \xstupleII {x=a[\xrange y1yn]}{\xD} }
{ \xstupleII {x=a[\xrange y1yn]}{\xD, y_k \in D_{y_k}\xdiff\{b\}} }
\quad\xcond{C$_y$}
\]
\[
\hspace{\xcondindent}
\xcond{C$_y$}:\qquad
	D_x \, \xintersect \,
	\left(\xUnion_{b_i\,\in\,D_{y_i},\;b_k=b} D_{a[\xrange b1bn]}\right)
	\, = \, \emptyset
\]
\[
\xRara\qquad
\frac
{ \xstupleII {x=a[\xrange y1yn]}
	{\xD} }
{ \xstupleII {x=a[\xrange y1yn]}
	{\xD, a[\xrange b1bn] \in D_{a[\xrange b1bn]} \xintersect D_x} }
\quad\xcond{C$_a$}
\]
\[
\hspace{\xcondindent}\xcond{C$_a$}:\qquad
D_{y_1} \times \ldots \times D_{y_n} = \{(\xrange b1bn)\}
\]
\resetlength{\mathindent}
\qed
\end{theorem}

\begin{proof}\quad
\paragraph{$(\Leftarrow)$}
Suppose $C \equiv \xtupleI {x=a[\xrangen y]}$ is closed under \xRSarr.
Then all values in domains of variables in $C$ are supported.

\noindent
(1) Take some $d \in D_x$.  $C$ is closed under \xRarx, thus also
$d \in \left(\xUnion_{b_i\in D_{y_i}} D_{a[\xrangen b]}\right)$.
Then there exists some $(\xrangen b)$ with $d \in D_{a[\xrangen b]}$.
This index and \xsubst {a[\xrangen b]}d support \xsubst xd.

\noindent
(2) For some $b \in D_{y_k}$ consider the necessarily failing condition of
\xRary. Thus a value $d$ exists in both $D_x$ and $D_{a[\xrangen b]}$,
for some $(\xrangen b)$ with $b_k=b$.  Assigning the $b_i$ to the $y_i$,
and \xsubst xd and \xsubst {a[\xrangen b]}d, is a solution supporting $b$.

\noindent
(3) Consider a value $d \in D_{a[\xrangen b]}$, and the following cases:
\\
(3.1) $(\xrangen b) \notin D_{y_1} \times \ldots \times D_{y_n}$
\\
The index $(\xrangen y)$ can not select the variable $a[\xrangen b]$;
however, $C$ remains satisfiable.  Therefore, there is a solution for $C$
that is indifferent to the value of $a[\xrangen b]$, and so supports
\xsubst {a[\xrangen b]}d.

\noindent
(3.2) $(\xrangen b) \in D_{y_1} \times \ldots \times D_{y_n}$
\\
(3.2.1) $\{(\xrangen b)\} = D_{y_1} \times \ldots \times D_{y_n}$
\\
Here the condition of \xRara\ is fulfilled, its consequence holds
and with it $d \in D_x$.  A supporting solution is therefore
\xsubst xd, \xsubst {a[\xrangen b]}d, \xsubst{y_i}{b_i} for all $i$.

\noindent
(3.2.2) some $D_{y_k}$ contains more than one element
\\
Consider some index $(\xrangen {b'})$ with $b'_k \not= b_k$
that also fulfills $D_x \xintersect D_{a[\xrangen {b'}]} \not= \emptyset$.
Such an index exists because otherwise \xRary\ would be applicable.
Choose a $d' \in D_x$ and instantiate
\xsubst x{d'}, \xsubst {a[\xrangen {b'}]}{d'}, \xsubst{y_i}{b'_i} for all $i$.
This solution to $C$ does not assign to $a[\xrangen b]$ and hence
supports \xsubst {a[\xrangen b]}d.

\paragraph{$(\Rightarrow)$}
Suppose here that $C$ is not closed under \xRSarr.
Then domains of some variables in $C$ contain unsupported values.

\noindent
(1) Assume \xRarx\ is applicable, that is,
$D_x \supset \xUnion_{b_i\in D_{y_i}} D_{a[\xrangen b]}$.
Then there is some value $d \in D_x, d \notin D_{a[\xrangen b]}$
for all $(\xrangen b) \in D_{y_1} \times \ldots \times D_{y_n}$.
Clearly, $d$ is not part of any solution.

\noindent
(2) Suppose some $b_k \in D_{y_k}$ could be removed by \xRary.
From the condition of \xRary\ it follows that with \xsubst {y_k}{b_k}
no index $(\xrangen b)$ can be found that allows the same value for
$x$ and $a[\xrangen b]$.

\noindent
(3) For a singleton index domain and so a possible application of \xRara\
consider \xsubst {a[\xrangen b]}d with $d \notin D_x$.
Such a $d$ can not be supported by $x$.
\qed
\end{proof}

\paragraph{Linearity requirement.}
It is necessary to restrict occurrences of variables.  Consider
the array $\mathit{xor} =
\{\xsubst {(0,0)}0, \{\xsubst {(0,1)}1, \{\xsubst {(1,0)}1, \{\xsubst {(1,1)}0 \}$
and the CSP
$\xP \equiv \xtupleII {0 = \mathit{xor}[y, y]}{\{y\in\{0,1\}\}}$.
\xP\ is inconsistent but stable under \xRSarr.

\paragraph{Origin of \xRSarr.}
Each rule in \xRSarr\ can be derived as an instance
of the general rule \xRac\ in Lemma~\ref{thm:acrule}.
Such a derivation, perhaps unsurprisingly, proceeds
along the same case distinctions
as in the $(\Leftarrow)$ part of the above proof.
We believe the derivation to be interesting in its own right,
but choose here the proof for its relative brevity.

\subsection{Arc-consistency and Compound Expressions}

The following result allows decomposition of
nested array expressions and equality constraints
for the purpose of establishing arc-consistency.
Expressions such as \mbox{$l[w, p] = l[w', p']$}
from the crossword example are decomposed with a fresh
variable into $v = l[w, p]$ and $v = l[w', p']$,
upon which arc-consistency can be enforced independently.

\begin{lemma}\label{thm:acdecomp}
Assume $C_t \equiv \xtupleI {s=t(v)}$ and $C_v \equiv \xtupleI {v=r}$
be linear constraints on, apart from $v$, distinct sets of variables.
The constraint $C \equiv \xtupleI {s=t\xsubsts vr}$ is arc-consistent
if  $C_t$ and $C_v$ are.
\qed
\end{lemma}

\begin{proof}
Suppose $C_t$ and $C_v$ are arc-consistent.

Any solution for $C_t$ assigns a value to $v$ that is also
supported by a solution to $C_v$, and vice versa.
Due to the conditions on variables, such solutions
do not assign to the same variables.  Therefore there
union is also a solution for $C$.
Thus, a supporting solution for any domain value of a variable
in $C_t,C_v$, and $C$, can be extended to a supporting solution for $C$.

Hence, $C$ is arc-consistent.
\qed
\end{proof}

\subsection{Domain Reduction and Transformation}\label{sec:redtrans}

As instances of \xRac, the rules in \xRSarr\ are
domain reduction rules by type.
From a semantical, and particularly from an operational, point of view,
however,
it may be worth to have instead transformation rules
which change the representation of constraints.

Consider \xRara, which applies if the index 
is fully instantiated.  That means 
we can also dispense entirely with the array look-up:
no choice is left.
The array expression can be replaced by the selected variable.
An alternative to \xRara\ would thus be
\[
\frac
{ \xstupleII {x=a[\xrangen y]}
	{\xD} }
{ \xstupleII {x=a[\xrange b1bn]}
	{\xD, a[\xrangen b] \in D_{a[\xrangen b]} \xintersect D_x} }
\quad\xcond{C$_a$}
\]
This rule is now both a transformation rule and a domain reduction rule.
Note that the domain reduction takes place between variables.
In presence of rules for primitive equality constraints
\xtuple{x=y} one can simplify even more
into a pure transformation rule:
\[
\xRarap\qquad
\frac
{ \xstupleII {x=a[\xrangen y]}{\xD}}
{ \xstupleII {x=a[\xrangen b]}{\xD}}
\quad\xcond{C$_a$}
\]
The combination of \xRarap\ and rules for \xtuple{x=y}
is equivalent to \xRara.


\section{A Non-naive Algorithm}

An exhaustive application of \xRSarr\ 
is computationally expensive, in part unavoidable due to the
strength of arc-consistency, and the large number of
variables involved in array constraints.
An inefficiency that can be remedied
is the large number of set operations
on domains,
due to fact that individual array variable domains
$D_{a[\xrangen b]}$ are read and processed multiple times.

The algorithm \xarrac\ (Fig.~\ref{fig:arrac}) reads every $D_{a[\xrangen b]}$
addressable by $(\xrangen y)$ at most once.
Consider $T = D_x \xintersect D_{a[\xrangen b]}$
for some $(\xrangen b) \in D_{y_1} \times \ldots \times D_{y_n}$.
$T$ is a subset of the intersection in the conclusion of \xRarx,
so it is necessarily part of the new domain of $x$, and only $D_x \xdiff T$
instead of $D_x$ needs to be subjected to further restriction.
With regard to \xRary, a nonempty $T$ implies that the side condition fails.
Thus, no $b_k$ of $(\xrangen b)$ can be removed from $D_{y_k}$ by \xRary.

\begin{figure}[htbp]
\setlength{\xalgotabinitial}{4mm}
\begin{algorithm}
\xafor\ all $i$: \hspace{\xalgokeyword} $B_i := D_{y_i}$
	\>\>\>\>\>\xalgocomment{index}\\
\xafor\ all $i$: \hspace{\xalgokeyword} $Y_i := D_{y_i}$
	\>\>\>\>\>\xalgocomment{potentially redundant for $y_i$}\\
$X := D_x$
	\>\>\>\>\>\xalgocomment{potentially redundant for $x$}\\
$S := \emptyset$
	\>\>\>\>\>\xalgocomment{indices skipped for $X$}\\[\xalgoskip]

\xawhile\ $B \not= \emptyset$ and some $Y_k \not= \emptyset$
	\>\>\>\>\>\xalgocomment{loop for $Y$ and $X$}\+\\
choose and remove $(\xrangen b)$ from $B$\\
\xaif\ some $b_k \in Y_k$
\xathen\+\\
$T := D_x \xintersect D_{a[\xrangen b]}$\\
\xaif\  $T \not= \emptyset$
\xathen\+\\
\xafor\ all $i$: \hspace{\xalgokeyword} $Y_i := Y_i \xdiff \{b_i\}$\\
$X := X \xdiff T$\-\-\\
\xaelse\ $S := S \xunion \{(\xrangen b)\}$\-\\[\xalgoskip]

\xawhile\+ $S \not= \emptyset$ and $X \not= \emptyset$
	\>\>\>\>\>\xalgocomment{rest loop for $X$}\\
choose and remove $(\xrangen b)$ from $S$\\
$X := X \xdiff D_{a[\xrangen b]}$\-\\[\xalgoskip]

\xafor\ all $i$: \hspace{\xalgokeyword} $D_{y_i} := D_{y_i} - Y_i$
	\>\>\>\>\>\xalgocomment{remove redundant values}\\
$D_x := D_x - X$
\end{algorithm}
\caption{\xarrac\ (core)}
\label{fig:arrac}
\end{figure}

Note that \xarrac\ makes a positive guess whether
values are supported.  If indeed in the end some domain is reduced
then \xarrac\ needs to repeat the run.
Indeed, if at some before the regular end of the run as described
in Fig.~\ref{fig:arrac} it is definite that some domain $D_{y_j}$
will be reduced, the run could terminate immediately, commit
the change to $D_{y_j}$, and restart.

The core part of \xarrac\ can itself be regarded as a complex domain
reduction rule, encapsulating \xRarx\ and \xRary.
The rule set $\{\mbox{\xarrac:core}, \xRara\}$ establishes arc-consistency.

\newcommand{\xl}[1]	{\mbox{\sffamily #1}}

\begin{example*}
Consider 
\mbox{$x\in\{\xl B, \xl C, \xl D\},y_1\in\{1,2\},y_2\in\{1,2,3\}$}
and \xtupleI{x=a[y_1, y_2]} and let $a$ be defined as
the array of constants
\begin{center}
\begin{tabular}
{@{\quad}c@{\quad}|@{\quad}c@{\quad}|@{\quad}c@{\quad}|@{\quad}c@{\quad}}
$(y_1,y_2)$  & 1 & 2 & 3 \\\hline
1 & $\xl A$ & $\xl B$ & $\xl C$ \\
2 & $\xl D$ & $\xl E$ & $\xl F$
\end{tabular}
\end{center}
The constraint is arc-consistent,
which \xarrac\ verifies as follows.
First it reads $a[1,1]=\xl A$.  Nothing is done.
It follows $a[1,2]=\xl B$. \xl B is in $D_x$,
so \xl B is a supported value for $x$,
and $1$ is supported for $y_1$ and $2$ for $y_2$.
The next step is reading $a[1,3]=\xl C$. This supports \xl C for $x$
and  $3$ for $y_2$.
Finally, $a[2,1]=\xl D$ is reached.  This supports the last missing value \xl D
for $x$, and moreover $2$ for $y_1$ and $1$ for $y_2$.

Support for all values in the domains was found, hence
arc-consistency is established.
Only one incomplete run was necessary,
skipping the indices $(2,2), (2,3)$ that are still permissible
by $(y_1,y_2)$.
\qed
\end{example*}

For one run of \xarrac\ (and ignoring $X$ here),
the number of iterations that enter the computation of $T$
has an upper bound of \xOrd {d^n} with $d$ the maximal size of
the domains of the $y_i$.
This reflects the number of possible different indices $(\xrangen b)$.
The lower bound, on the other hand, is only \xOrd {d}.
It is reached when every iteration reduces all (nonempty) $Y_i$
by an element, and occurs if the constraint is arc-consistent
and every instantiation of $(\xrangen y)$ is part of a solution.

An operationally useful side effect of \xarrac\ is that
it can also yield the variables that contain the supporting values.
Initially, all variables $a[\xrangen b]$
are part of the constraint, whereas after complete instantiation
of the index $(\xrangen y)$ only the variable
$a[\xrangen y]$ is constrained and contains support.
\xarrac\ regards those variables $a[\xrangen b]$ as supporting,
for which the intersection $T$ is nonempty.

\xarrac\ was implemented in
\xeclipse\ (\cite{www:eclipse}),
using the finite domain primitives of \texttt{lib(fd)}.
An implementation of \xRSarr\ in the same environment
was compared to \xarrac\ by testing it against an instance
of the crossword problem, and was roughly 50\% slower.


\section{Final Remarks}

\subsection{Related Work}

The established precursor of array constraints is
the \texttt{element} constraint
(\cite{dh:chip}).
It is the one-dimensional specialisation, and usually
the look-up list that links index and result
is restricted to consisting of constants.

Arrays in \xopl\ (\cite{pvh:opl,www:oplst}) are similarly
general as in this work.
In \cite{mh:oplpl} on \xopl++ a model of the stable marriages
problem is described that employs an array of variables
indexed by a variable.
Constraint propagation of array expressions in \xopl\
is strictly weaker, however.  For all three cases treated by \xRSarr\
we could construct simple examples using small 2-dimensional arrays
in which reduction of domains is possible but not performed,
see Figures~\ref{fig:oplxy} and \ref{fig:opla}.

\cite{chj:aklfd} describes an implementation
of \texttt{element} 
using indexicals in AKL(FD),
in which the look-up list can consist
of domain variables.  It is equivalent to a one-dimensional
instance of \xRSarr.

In \cite{nb:gcgpsnec} a new constraint \texttt{case} is
proposed that subsumes
multidimensional array constraints with arrays of constants.
An algorithm which seems similar in effect to using \xRSarr,
based on graph theory, is outlined.

\cite{hotk:suocsm}, on unifying optimisation and constraint satisfaction
methods,
studies a continuous relaxation of \texttt{element}
with a look-up list of variables with continuous domains
by using a cutting-planes approach.

\subsection{Conclusions}

We study here the use of arrays in constraint programming
mainly from a theoretical point of view.  There are good arguments
suggesting that arrays are beneficial in constraint models, however.
Indices on objects are basic in mathematics.
\texttt{element} is implemented in many constraint systems.
Arrays with multiple dimensions have long been used
in imperative, now object-oriented, languages.
These language styles obviously inspired
\xopl\ (\cite{pvh:opl}, more so \cite{mh:oplpl}),
a successful constraint programming system.
Yet it would be desirable to have large examples of uses of
multidimensional arrays.

Such problems could also be used to evaluate the use
of arc-consistency as the objective in constraint propagation.
It is now clear from practical experience that
the notion of consistency that is most advantageous depends
on the problem.  Sometimes a weaker notion such as
bound or range consistency might suffice, for example
applied in the early stages of solving 
and later replaced by full arc-consistency.
\xRSarr\ provides a starting point
for obtaining reduction rules for those consistency notions,
which are subsumed by, yet very similar to, arc-consistency.


\section{Acknowledgement}

Krzysztof Apt suggested the topic of this work and made many helpful
comments. I am thankful also to referee comments 
on an earlier presentation of the subject.


\bibliographystyle{abbrv}
\bibliography{bibfile}


\begin{figure}[htbp]
\begin{small}
\begin{verbatim}
enum      Dz  { i, j    };
enum      Dy  { k, l, m };
enum      Da  { p, q, r };

Da        a[Dz, Dy] = #[  i:  #[k:p, l:q, m:r]#,
                          j:  #[k:p, l:q, m:r]#   ]#;
var       Da  x;
var       Dz  z;
var       Dy  y;
var       Dz  u;
var       Dy  v;

solve {   v <> l;         //              OPL       arc-consistency
          a[u, v] = x;    //  x  in   { p, q, r }      { p, r }
                          //
          a[z, y] = q;    //  y  in   { k, l, m }       { l }
};
\end{verbatim}

\end{small}
\caption{\xopl: non-applied \xRarx, \xRary}
\label{fig:oplxy}
\end{figure}

\begin{figure}[htbp]
\begin{small}
\begin{verbatim}
enum      Dy  { i, j, k };
enum      Da  { p, q, r };

var       Da  a[Dy];
var       Da  x;
var       Dy  y;

solve {   y = j;
          x <> q;       //                OPL       arc-consistency
          x = a[y];     //  a[j]  in  { p, q, r }      { p, r }
};
\end{verbatim}
\end{small}
\caption{\xopl: non-applied \xRara}
\label{fig:opla}
\end{figure}


\end{document}